\documentclass[12pt]{article}
\usepackage{amsmath,amssymb,graphicx,mathrsfs,hyperref}

\newcommand{\be}{\begin{equation}}
\newcommand{\ee}{\end{equation}}
\newcommand{\bea}{\begin{eqnarray}}
\newcommand{\eea}{\end{eqnarray}}

\newcommand{\wh}{\widehat}

\newcommand{\sk}{k^{\frac{1}{2}}}
\newcommand{\cc}{\frac{{\cal C}}{\hbar}}
\newcommand{\ccc}{{\cal C}/\hbar}
\newcommand{\cccc}{\frac{\hbar}{{\cal C}}}
\def\({\left(} \def\){\right)}

\begin{document}
%%%%%%%%%%%%%%%%%%%%%%%%%%%%%%%%%%%%%%%%%

\title{\vspace{-2.1in}
%\begin{flushright} {\footnotesize CERN-PH-TH/2011-204}  \end{flushright}
%\vspace{3mm}
\vspace{0.3cm} {Semiclassical black holes expose forbidden charges and censor divergent densities }}
\author{\large Ram Brustein${}^{(1)}$,  A.J.M. Medved${}^{(2)}$ \\
 \hspace{-1.5in} \vbox{
 \begin{flushleft}
  $^{\textrm{\normalsize
(1)\ Department of Physics, Ben-Gurion University,
    Beer-Sheva 84105, Israel}}$
$^{\textrm{\normalsize (2)  Department of Physics \& Electronics, Rhodes University,
  Grahamstown 6140, South Africa }}$
 \\ \small \hspace{1.7in}
    ramyb@bgu.ac.il,\  j.medved@ru.ac.za
\end{flushleft}
}}
\date{}
\maketitle
%%%
\vspace{-.4in}
\begin{abstract}
Classically, the black hole (BH) horizon is a rigid surface of infinite redshift; whereas the uncertainty principle dictates that the semiclassical (would-be) horizon cannot be fixed in space nor can it exhibit any divergences. We propose that this distinction underlies the BH information-loss paradox, the apparent absence of BH hair, the so-called trans-Planckian problem and the recent ``firewall'' controversy. We argue that the correct prescription is to first integrate out the fluctuations of the background geometry and only then evaluate matter observables. The basic idea is illustrated using a system of two strongly coupled harmonic oscillators, with the heavier oscillator representing the background. We then apply our proposal to matter fields near a BH horizon, initially treating the matter fields as classical and the background as semiclassical. In this case, the average value of the associated current does not vanish; so that it is possible, in principle, to measure the global charge of the BH. Then the  matter is, in addition to the background, treated  quantum mechanically. We show that the average energy density of matter as seen by an asymptotic observer is finite and proportional to the BH entropy, rather than divergent. We discuss the implications of our results for the various controversial issues concerning BH physics.

\end{abstract}
%\maketitle***
\newpage

\section{Introduction}

Quantum field theory in curved spacetime provides many useful  methods for learning about cosmological and black hole (BH) physics \cite{BD,SF,RW}. It  was hoped, at one time, that studies of this nature could provide at least a starting point toward a full theory of quantum gravity. It has, however, since been accepted that this theory can only effectively describe gravitating systems in certain well-controlled settings.

In a BH context, one is treating the background spacetime  as a classical entity while treating the matter as quantum fields. This caveat is well understood. A case in point is Hawking's famous demonstration that BHs radiate thermally \cite{Hawk}. Much attention is given there, as well as in subsequent related studies, to keeping the quantum field fluctuations under control. Otherwise, the back-reaction from the fluctuations would threaten to distort the background geometry and undermine the legitimacy of the calculation.

How does one operationally keep the back-reaction in check? The usual means is to take the system of interest to be sufficiently massive or arrange the gravitational coupling to be sufficiently small, usually both. Then, as long as the quantum fluctuations of the background are suppressed in comparison to the quantum fluctuations of the matter fields, one expects that the back-reaction  effects have been rendered harmless.

The implicit assumption that underlies this reasoning is that the effects of the quantum fluctuations of the background on the fields are less important. This is a reasonable assumption because the quantum wavelength of the matter is much larger than that of the background. However, this implicit assumption turns out to be incorrect in some important cases \cite{RB,Dvali1,Veneziano}. This is the starting point of the current investigation.

For concreteness, let us discuss the Schwarzschild BH. In this case, the classical limit can be expressed in terms of a ``classicality'' parameter $C_{BH}$ of a BH with mass $M_{BH}$ \cite{RB}. This parameter is defined in terms of the ratio of the Compton wavelength of the BH
$\;\lambda_{BH}= \hbar/M_{BH}\;$ to its Schwarzschild radius
$\;R_S= 2 G M_{BH}\;$. Using the relation between Newton's constant and the 4D Planck length $\; l_p=\sqrt{\hbar G}\;$, one finds that the ratio is given by
\begin{equation}
C_{BH}\;=\;\frac{\lambda_{BH}}{R_S}\;=\;2 \frac{l_p^2}{R_S^2}\;\sim\; \frac{1}{S_{BH}} \;,
\label{class}
\end{equation}
where $S_{BH}$ is the Bekenstein--Hawking entropy, the speed of light has been
set to unity and, for later convenience, we  regard
$\;C_{BH}=S^{-1}_{BH}\;$ as  an identity.
For a macroscopic BH, the value of $C_{BH}$ is extremely small. The parameter $C_{BH}$ is called $1/N$  by Dvali and Gomez \cite{Dvali1} and serves a similar purpose.

The classical limit is defined as the limit when $\;C_{BH}\to 0\;$ for a fixed $R_S$. One can also think about the classical limit as  $\;``M_{BH}\to\infty"\;$, $\;``G\to 0"\;$ and $G M_{BH}$ fixed,
this being  the limit for which the theory of quantum fields in curved space applies.  Under these conditions,
the background  can be safely declared as fixed. This could be viewed, more generally, as the definition of a classical background, when the
metric $g_{\mu\nu}$ can be regarded as frozen  at some classical value $(g_{\mu\nu})_c$ that is determined by the classical Einstein field equations.

The issue that we address in this paper is whether corrections to this limit are important and in which cases. Following \cite{RB}, we argue that they could be. Generically, for small $C_{BH}$, we expect power corrections in $C_{BH}$. If one calculates a quantity that has some finite value in the classical limit, the corrections will be small and irrelevant. However, if some expectation values vanish or diverge in this limit, then the corrections can be very significant. Our
main impetus being that
the semiclassical regime is still a quantum one. A quantum background will fluctuate and, even if the back-reaction of the matter fluctuations on the massive background is negligible, who is to say the converse is true?

This last point has become especially pertinent in light of some recent observations regarding semiclassical BHs \cite{RM,RB}. The notion of a classical BH solution with a fixed horizon representing a surface of infinite redshift is realized  only in the strict classical limit $\;C_{BH}=0\;$ and does not survive in the semiclassical picture. The more accurate description is in fact a superposition of quantum wavefunctions, none of which could on its own  describe a classical spacetime; also see \cite{englert}. A similar idea had previously been put forward in the ``fuzzball'' description  of BHs   \cite{FB1,FB2}. In this
description, the BH is an ensemble of microstates.

And yet, shouldn't the semiclassical BH  wavefunction be able to mimic the properties  of a classical BH? The answer is that it might or might not depending on the questions being posed. For example, BH  thermodynamics seems to apply to any BH  with sufficient mass to radiate thermally, whereas the information  paradox \cite{info1,info2,info3,info4}, the trans-Planckian problem \cite{trans1,trans2,trans3} and the recently posed ``firewall'' argument \cite{AMPS} are strong indicators that something indeed went wrong with the naive semiclassical interpretation. Briefly, the information  paradox asks what happened to the information stored inside of a BH  after it evaporates. Apparently, an initially pure state would have transitioned into a final state that is mixed. The trans-Planckian problem is that the process of Hawking radiation involves exponentially large energy scales unless a cutoff is imposed. But, on the other hand, such a cutoff would  jeopardize the thermality of the spectrum. The firewall argument is basically a modern reinterpretation of the information paradox, which we address later in the paper.

That the BH  horizon cannot survive  in even a highly semiclassical world, $\;C_{BH}\ll 1\;$, may seem like a rather strong assertion  but follows quite naturally from the quantum uncertainty principle \cite{RJ,RB}. If the horizon is fixed in space, then the BH must have infinite entropy and energy; hence, $C_{BH}=0$ and the BH  is classical by definition. Conversely, once $C_{BH}$ is finite, quantum effects will censor both the BH  singularity and the (no-longer-required) surface of infinite redshift. Meanwhile, the location of this ``would-be horizon'' can  no longer be viewed as  rigid; it  is free to fluctuate, just like any other quantum degree of freedom. Additional discussions  on the important distinction between finitely and infinitely massive BHs from a somewhat different point of view can be found in \cite{Dvali1,Dvali2}.

We propose that the background should, in all cases, be treated as a fluctuating entity. Matter-field expectation values should be evaluated for a finite $C_{BH}$  and, only then, should the background fluctuations be integrated out and the limit $\;C_{BH}\ll 1\;$ be invoked. This proposal is different than the usual convention of fixing the background metric, {\it a priori}, to an appropriate classical solution. There are likely many valid ways of accomplishing such a task, but we will  make the concrete proposal of representing the background fluctuations by a Gaussian wavefunction, as in \cite{RM} where it is justified. The exact nature of the BH wavefunction is not essential to the results that we obtain.

Our proposed procedure will be demonstrated explicitly for the simple case of a quantum harmonic oscillator (Subsection~2.1).  It will turn out that we are actually coupling the original oscillator to a (background) heavy one and then integrating out the heavy oscillator in the standard way (Subsection~2.2). Interestingly, when adopting this interpretation, we find that the coupling between the two oscillators is necessarily strong. This interpretation also suggests a similarity between our model and that of Caldeira and Leggett \cite{CL}, which we are able to exploit (Subsection~2.3). Another similar thought experiment with a light and heavy string was proposed in \cite{Veneziano}. The importance of our methodology when one encounters classical zeroes and infinities is also made clear
(Subsection~2.4).

When regarded as a fixed surface of infinite redshift, a BH  horizon can  be shown to induce an infinitely strong coupling for the matter fields (Subsection~3.1). We  find, on the other hand, that  integrating out the fluctuations of the (would-be) horizon can lead to a picture that differs substantially from this classical one ---  irrespective of  the matter fields being
classical  (Subsection~3.2) or quantum (Subsections~3.3,~3.4).
Our main assertion is  that this perspective is both necessary and sufficient
for  resolving the aforementioned conundrums (Subsection~3.5, Section~4). Indeed, there is already a tradition of using horizon fluctuations to address the trans-Planckian problem of Hawking radiation \cite{fluc1,fluc2,fluc3,fluc4,fluc5}.

We typically  work in units in which  fundamental constants
(besides $\hbar$ and $G$) are set to unity,  all oscillators have their masses fixed to unity (or absorbed into their spring constants) and normal ordering of their operators is consistently applied ({\it i.e.}, the ground states are calibrated to zero energy).  Calculations relevant to Subsection~2.2 and
Subsection~3.3 are
deferred to  Appendices~A and B respectively.

\section{Examples with oscillators}

The basic principles of our proposal can be illustrated for the simplest of systems, a quantum harmonic oscillator.  We will start by discussing a harmonic oscillator with a fluctuating spring constant and show that physical quantities receive corrections going as a power series in the strength of
fluctuations or, equivalently, in the relevant classicality parameter.
It is next  shown
that the fluctuating spring constant can be induced by strongly
coupling
the first oscillator
 to a second, heavier one. Then, after discussing the analogy between
this model and that of Caldeira and Leggett, we consider classically vanishing
 and divergent  interaction terms between the two oscillators and show that
 the quantum fluctuations of the heavy oscillator render them finite.

Since we will use the quantum harmonic oscillator throughout this section, let us recall its Hamiltonian
$\; \wh{H}  = \frac{1}{2}\wh{p}^2+\frac{1}{2}k_c \wh{x}^2\;$
or, in terms of creation and annihilation operators $a^\dagger$ and $a$,
$\;\wh{H} = \hbar\sk_c  a^{\dagger} a \;$,
where $k_c$ denotes the ``classical'' spring constant.

Let us further consider some arbitrary quantum state for this system,
$\;|\psi \rangle = \sum\limits_{n=0}^{\infty} c_n |n\rangle \;,$
where the $c_n$'s are complex numbers satisfying
$\;\sum\limits_{n} |c_n|^2=1\;$ and
$\;\wh{H} |n\rangle=\hbar\sk_{c}n|n\rangle\;$.
The energy $\;E=\langle \psi | \wh{H} | \psi \rangle\;$ goes as
\be
E\;=\;\hbar\sk_{c} \sum_{n} n|c_n|^2\;.
\label{energy}
\ee
Here, $k_c$ is a given fixed quantity.

\subsection{A quantum harmonic oscillator with a fluctuating spring constant}

In reality, everything is quantum and so everything can fluctuate. For the oscillator, we will  assume that a fluctuating spring constant can be used to model the quantum effects of the neglected degrees of freedom. And so we replace the constant $k_c$ with a fluctuating variable $k$ and assume, for simplicity, that its probability distribution is a Gaussian centered about the classical value $k_c$ and that it is restricted to positive values,
\be
\chi [k;k_c, {\cal C}] \;=\; {\cal N}^{-1/2}\; e^{-\frac{\cal C}{2 \hbar} (k-k_c)^2}\Theta(k) \;.
\ee
Here, ${\cal C}$ is a dimensional parameter that controls the width of the
Gaussian and
\be
{\cal N}\;=\; \frac{1}{2}\left(\frac{\pi\hbar}{{\cal C}}\right)^{\frac{1}{2}}\left[1+{\rm erf}\left(\sqrt{\frac{{\cal C}}{\hbar}}k_c\right)\right].
\ee
is a normalization factor that  guarantees  $\;\int\limits^{\infty}_0 dk\;
\chi^2[k]=1\;$.

The ratio $\hbar/{\cal C}$ determines the strength of the fluctuations in $k$,
$\;\Delta k^2\sim \hbar/{\cal C}\;$. Since the context of this discussion presumes that $\;k_c\gg |\Delta k|\;$,
\be
\frac{ {\cal C} k_c^2}{\hbar}\;\gg \; 1 \;.
\label{large}
\ee
We see that the parameter
\be
C_{OSC}=\frac{\hbar}{ {\cal C} k_c^2}
\ee
is the classicality parameter of the oscillator.
A truly classical background corresponding to a fixed spring constant is only attained in the limit $C_{OSC}\to 0\;$, in which case $\;\chi^2=\delta(k-k_c)\;$. When $\;C_{OSC} > 1\;$, then the fluctuations in the spring constant are comparable to its average value. For such a case, the background is intrinsically quantum.

Now, to recalculate the energy, we  follow the standard practice of  integrating out the background mode.
This means performing the integral $\;{\cal I}=\int dk \; k^{1/2} \chi^2[k] \;$:
\bea
{\cal I} &=&{\cal N}^{-1}\int\limits^{\infty}_0 dk\; k^{1/2}\; e^{-\cc (k-k_c)^2} \nonumber \\
&=& \sk_c+\sqrt{\cccc}{\cal N}^{-1}\sk_c\int\limits^{\infty}_{-\sqrt{\ccc}k_c} dl\;
\left[\frac{1}{2}\sqrt{\cccc}\frac{l}{k_c}-\frac{1}{8}\cccc\frac{l^2}{k_c^2}
+\cdots\right]\; e^{-l^2}\;,\nonumber \\
&&
\eea
with the integrated term in the last line following from the normalization of $\chi$ and the
ellipsis  denoting the higher-order terms from the Taylor expansion of
the square root.

The term linear in $l$ in the expansion,
\bea
\int\limits^{\infty}_{-\sqrt{\ccc} k_c} dl\; l \;e^{-l^2}
&=&\frac{1}{2} e^{-\cc k_c^2}\;,
\label{expdec}
\eea
is exponentially small, as are all the odd terms in the series.
For the quadratic term,
$\;\int dl\; l^2\; e^{-l^2}=\frac{1}{2}\int dl\; e^{-l^2}\;$, so that
$\;
{\cal I}\;=\; \sk_c\Big(1-\frac{1}{16}C_{OSC}+\cdots
\Big)\;,
$
where $\cdots$ stand for higher powers of $C_{OSC}$.
The energy is corrected in a similar way,
\be
E\;=\;E_c \;
\Big(1-\frac{1}{16} C_{OSC}\;+\cdots\Big) \;,
\label{correct}
\ee
with $E_c$ denoting the previous classical value.

The conclusion is that, for small $C_{OSC}$, the energy $E$ receives power corrections in $C_{OSC}$. Since $C_{OSC}$ is small, the corrections do not change the classical value in a particularly significant way.

\subsection{A fluctuating spring constant from a strongly coupled heavy oscillator}

We can  obtain  the same basic picture as in Subsection~2.1 by introducing the background  into
the theory as a second, heavier oscillator. Let us begin with the following Hamiltonian for a coupled-oscillator system:
\be
\wh{H} \;= \;\frac{1}{2}\wh{p}_x^2+\frac{1}{2}\omega_x^2 \wh{x}^2 +\frac{1}{2}\wh{p}_y^2 +
\frac{1}{2}\omega_y^2\left(\wh{y}-y_0\right)^2 +\frac{1}{2}\wh{y}\wh{x}^2\;.
\ee
The interaction term $\wh{y}\wh{x}^2$ contains a dimensional constant that has been set to unity. The $y$ oscillator is taken to be much ``heavier" than the $x$ oscillator, $\;\omega_y^2\gg \omega_x^2\;$. We say ``heavy" because the frequency of the oscillator is the analogue of the mass term for a relativistic field.

When the two oscillators are weakly coupled $\;y_0\ll\omega_x^2\;$, then the energy of the $x$ oscillator can be evaluated using perturbation theory. In this case, with the heavy oscillator assumed to be in its ground state, the relative change in the energy is $\frac{1}{2} y_0/ \omega_x^2$.

But the strong-coupling limit $\;y_0 \gg \omega_x^2 \;$ is more relevant to the
upcoming BH discussion. In this case, the ``normal" potential term for the $x$ oscillator can be disregarded. Let
us  relabel some of  the operators and parameters: $\;\wh{y}\to \wh{k}\;$,
 $\;y_0\to k_c\;$, $\;\wh{p}_y\to \wh{p}_k\;$ and $\;\omega_y\to {\cal C}\;$.
The Hamiltonian then becomes
\be
\wh{H} \;=\; \frac{1}{2}\wh{p}_x^2+\frac{1}{2}\wh{k} \wh{x}^2 +\frac{1}{2}\wh{p}_k^2 +
\frac{1}{2}{\cal C}^2(\wh{k}-k_c)^2 \;.
\label{hamil}
\ee

In the strong-coupling regime, $\wh{k}$ effectively plays the role of spring constant for the  light $x$ oscillator in addition to position for the  heavy background oscillator.
Meanwhile,  ${\cal C}^2$ serves as the spring constant for the heavy
oscillator. To complete the analogy with Subsection~2.1,
we restrict the operator $\wh{k}$  to positive eigenvalues as a boundary condition.

Let us once more rewrite the Hamiltonian~(\ref{hamil}) as
\be
\wh{H}\;=\;\frac{1}{2}\wh{p}_x^2+\frac{1}{2}k_c \wh{x}^2 +\frac{1}{2}\wh{p}_k^2 +\frac{1}{2}{\cal C}^2(\wh{k}-k_c)^2
+\frac{1}{2}(\wh{k}-k_c)\wh{x}^2\;
\label{hammy}
\ee
and then regard the last term $\frac{1}{2}(\wh{k}-k_c)\wh{x}^2$ as the interaction term in the Hamiltonian.

In general, we need to consider wavefunctions of the heavy and light oscillators $|\psi_{k,x}\rangle$. However, for our purposes, it will suffice to assume from now on that the background oscillator is in its ground state. If, in addition,  we temporarily ignore the interaction term, then instead of the most general state $\;|\psi_{k,x}\rangle$ we need to only consider  $\;|\psi_{k,x}\rangle=|0_k\rangle|\psi_x\rangle\;$. This state can be explicitly written in the $k$ representation for the heavy oscillator and in the energy representation for the light oscillator,
\be
\langle k|\psi\rangle \;=\;{\cal N}^{-1/2}  e^{-\frac{1}{2}\cc(k-k_c)^2}
\Theta(k)\sum_{n=0}^{\infty} c_n |n\rangle \;.
\label{wavey}
\ee

We next want to evaluate the energy for this set-up,
using only the approximate wavefunction~(\ref{wavey}).
The justification for this, as well as for ignoring the back-reaction of the light oscillator on the heavy one, is provided in Appendix~A.

With the interaction term restored, the Hamiltonian can be re-expressed in a   way that is appropriate  for our choice of representations,
\be
\wh{H} \;=\; \hbar\sqrt{\wh{k}}  a^{\dagger} a + \frac{1}{2}\wh{p}_k^2 +\frac{1}{2}{\cal C}^2(\wh{k}-k_c)^2\;.
\ee
Since the background oscillator is in its ground state, the last two terms will make no contribution
and  can safely be ignored.

For simplicity, let us calculate $E_m=\langle m|\wh{H}|m\rangle$,
\bea
E_m &=& \hbar m{\cal N}^{-1}\int\limits_{0}^{\infty} dk \;\sk    \; e^{-\cc(k-k_c)^2} \;,
\eea
from which it is  evident that the corrections to the energy will be the same as those in Eq.~(\ref{correct}),
\be
E_m \;=\; (E_m)_c(1- 16 C_{OSC}+\cdots)\;,
\label{correct2}
\ee
where $\;(E_m)_c= \hbar\sqrt{k_c} m\;$ is the energy of the light oscillator when the heavy one is treated as classical.

\subsection{The Caldeira-Leggett Perspective \label{CL}}

Our coupled-oscillator system is reminiscent
 of the Caldeira--Leggett (CL) model \cite{CL},
\be
\wh{H}_{CL} \;=\; \frac{1}{2}\wh{p}_x^2+\frac{1}{2} k\wh{x}^2 +
 \frac{1}{2}\sum_i^N\left[\wh{p}^2_{y_i} +k_i \wh{y}_i^2\right]
+\wh{x}\sum_i^N c_i \wh{y}_i\;.
\ee
Here, the interaction with the background (or environment) is modeled with a collection of $N$  weakly coupled oscillators.
The CL  background is meant to represent a thermal bath of oscillators, so these are set in a mixed state;
 whereas our heavy oscillator has been set in a pure state, its ground state.
 Nevertheless, both models amount to a strong coupling between a simple system and its environment. So that, after
integrating out the respective backgrounds, one might anticipate some qualitatively similar results.

We can be more specific about this connection by inspecting
the results of \cite{RV}, where the late-time form of
the CL reduced density matrix has been obtained. From \cite{RV} it can be deduced that observables like the energy and entropy have exponential corrections away from their classical values in the small temperature limit  and power-law corrections in the limit of large temperature.

The expansion parameter is the classicality parameter of the CL model $\;C_{CL}=\hbar \omega /k_B T\;$, where $\omega$ is the frequency of the $x$-oscillator and $T$ is the temperature of the bath. The parameter $C_{CL}$ is then the ratio of a quantum energy for the system $\hbar\omega$ to a classical one for the background $k_B T$. This identification makes it clear that $C_{CL}$ is a dimensionless quantity that is tracking the classicality of the CL system. When $\;C_{CL}\ll 1 \;$, it means that the system is semiclassical, whereas  $\;C_{CL}\gg 1\;$
means it  is highly quantum.  In the latter case, the leading correction is of
order  $e^{-C_{CL}}$, but the corrections
for the former case  are a power series in $C_{CL}$.

The classicality parameter of our model $C_{OSC}$ is the analogue of the expansion parameter $C_{CL}$.  Up till now, we have been assuming the semiclassical regime of small $C_{OSC}$ and found that corrections go as a power series in $C_{OSC}$ ({\it cf}, Eq.~(\ref{correct})). Hence, our model agrees with that of CL  for  this region.

But what about when $C$ becomes large? Our choice of Gaussian wavefunction for the background is geared specifically towards the semiclassical limit.  We expect that the correct choice of wavefunction for the quantum regime cannot be a Gaussian and remains an interesting open question. The answer should be relevant to quantum description of small (Planck-sized) BHs, and we hope to address this scenario in a future article.

\subsection{Quantum fluctuations tame classically vanishing or divergent interactions}

We would now like to see  what happens for interactions between the heavy and light  oscillator  such that, classically, these either vanish or diverge.
This type of scenario will be directly relevant  to the discussion on
BHs.

For example, let us consider the following Hamiltonian:
\bea
\wh{H}&=&\frac{1}{2}\wh{p}_x^2+\frac{1}{2}k_c \wh{x}^2 +\frac{1}{2}\wh{p}_k^2 +\frac{1}{2}{\cal C}^2(\wh{k}-k_c)^2
+\frac{1}{2}(\wh{k}-k_c)\wh{x}^2 \cr &+& g_1 \frac{(\wh{k}-k_c)^2}{k_c^2}\; k_c\wh{x}^2+ g_2 \frac{k_c^2}{(\wh{k}-k_c)^2} k_c\wh{x}^2\;.
\label{hammyInt}
\eea
The new interaction term $\;H_I^{(1)}=g_1 \frac{(\wh{k}-k_c)^2}{k_c^2}\; k_c\wh{x}^2\;$ would vanish in the classical limit $C_{OSC}\to 0$. The other new
term $\;H_I^{(2)}=g_2 \frac{k_c^2}{(\wh{k}-k_c)^2} k_c\wh{x}^2\;$ diverges
for the same. We would like to know what their fate is when $C_{OSC}$ is
rather regarded as small but still finite.

We begin with $H_I^{(1)}$ and evaluate its expectation value with respect to
the heavy-oscillator wavefunction. To leading order,
\bea
H_I^{(1)}&=& g_1 k_c \wh{x}^2 {\cal N}^{-1}\;\int\limits^{\infty}_{0} dk \frac{(\wh{k}-k_c)^2}{k_c^2} e^{-\frac{\cal C}{ \hbar} (k-k_c)^2}
\nonumber \\  &=& \frac{g_1}{2}  k_c \wh{x}^2 \frac{\hbar}{{\cal C}k_c^2}
\;=\;\frac{g_1}{2}  k_c \wh{x}^2 C_{OSC} \;.
\eea
Clearly, a classically vanishing interaction term has been rendered finite and proportional to the small but finite classicality parameter $C_{OSC}$.

Let us now discuss $H_I^{(2)}$, which has an expectation value with the heavy-oscillator wavefunction of the form
\be
H_I^{(2)} =g_2 k_c \wh{x}^2 {\cal N}^{-1}\;\int\limits^{\infty}_{0} dk \frac{k_c^2}{({k}-k_c)^2} e^{-\frac{\cal C}{ \hbar} (k-k_c)^2} \;.
\label{HI2}
\ee

This integral obviously  diverges, as the integrand grows too strongly  as $\;k\to k_c\;$. Quantum theory, however, instructs us that the inverse operator $(\wh{k}-k_c)^{-2}$ is  only to  be defined after  removing the zero modes of $\;(\wh{k}-k_c)^{2}\;$. An elegant way of performing this removal is to evaluate Eq.~(\ref{HI2}) using  complex contour integration. The appropriate contour can be identified by recognizing that the integral in Eq.~(\ref{HI2}) is
of a form similar to  $\;\Gamma(z)=\int\limits_0^\infty dt\; e^{-t^2} t^{2z-1} \;$ when $\;z=-1/2\;$. Hence, one should choose the Hankel contour, as this is used in the standard definition of the $\Gamma(z)$ function for negative values of $z$.

With this choice of contour  and the realization that the resulting  integral
$\;I=\int\limits^{\infty}_{-C_{OSC}^{-1/2}}dl\;e^{-l^2}l^{-2}\;$
is, to a good approximation, $\;I=2\int\limits_{0}^{\infty} dt\; e^{-t^2} t^{-2} \;$, it follows that $\;I=2\Gamma(-1/2)=- 4 \sqrt{\pi}\;$. Collecting the various factors in Eq.~(\ref{HI2}), we have to leading order
\be
H_I^{(2)} =-4  g_2 k_c \wh{x}^2 \frac{1}{C_{OSC}}\;.
\ee
Thus, a classically positive and divergent interaction term has been rendered  finite and negative. It is proportional to a large but finite  quantity, the inverse of the classicality parameter $1/C_{OSC}$.

\section{Semiclassical Black Holes}

It is often said that nothing special happens at the horizon of a large BH, as a free-falling observer will pass right through  with no (immediate) physical consequences.~\footnote{This was, at least, the ``pre-firewall'' consensus of opinion.} On the other hand, it is indeed special to an asymptotic observer who sees strange things such as an apparently divergent redshift and a thermal bath with  an apparently infinite energy. Because of these dissenting perspectives, it is sometimes claimed that such infinities are a consequence of a using poorly selected  coordinates or making a misguided choice for the vacuum state. This is true to an extent (see further on), but it is also true that these  divergences lead to quandaries like the trans-Planckian problem, as  Hawking's calculation of  BH  radiation is necessarily carried out with an asymptotic observer in mind. Our main point is that these divergences should never have been  there in the first place, assuming that the BH is not infinitely massive.

Because of the infinite redshift at the horizon of a (classical) BH, conditions for  strong coupling between matter fields  and  the background are prevalent. We will be able to illustrate this below with a simple but explicit example of a static, spherically symmetric BH in Schwarzschild coordinates, with the classical position of the horizon specified by $\;r=R_c\;$. Recall  that $\;R_c = 2M_{BH}G\;$ is taken as finite and that the classical limit is defined by $\;C_{BH}\to 0\;$ or, equivalently,   $\;M_{BH}\to\infty\;$, $\;G\to 0\;$.

We will be using the wavefunction for the S-wave mode of a Schwarzschild BH in 4D Einstein gravity \cite{RB}.
Let us start from the wavefunction in the entropy representation  \cite{RM}. The wavefunction has to yield expectation values that reproduce the classical values in the classical limit, according to the Bohr correspondence principle. In this case, the average and variance of the entropy have to be recovered. Specifically, $\langle \wh{S}_W \rangle = A/ 4 l_p^2$, $A$ being the area of the horizon and $ \Delta {S}_W =  A/2 l_p^2$.
The simplest wavefunction that satisfies these two requirements was found in \cite{RM}, 
\begin{equation}
\Psi\left(S_W\right)\;\sim\; e^{\hbox{$-\frac{l_p^2}{2A} \bigl(S_W-A/4 l_p^2\bigr)^2$}}\;.
\label{swavefunction}
\end{equation}
Higher moments $\langle (\wh{S}_W)^n\rangle$ are not constrained by the simple reasoning and so could be modified.

Considering specifically the S-wave mode of a Schwarzschild BH in 4D Einstein gravity, $\;S_W= S_{BH}=\pi R_c^2/l_p^2\;$. In this case we may replace the entropy wavefunction $\Psi(S_W)$ by a radial wavefunction $\Psi(R)$. Given that $\;R^2\gg l_p^2\;$, the entropy wavefunction is highly peaked about $S_W=S_{BH}$ and so the radial wavefunction  is highly peaked about $R=R_c$. Then we could approximate the radial wavefunction as in \cite{RB}
\begin{eqnarray}
\label{solutionfin}
\Psi(R)\;=\;{\cal N}^{-1/2} e^{\hbox{$-\frac{\pi}{2 \hbar G} (R-R_c)^2$}}\;,
\end{eqnarray}
where $\;{\cal N}= 4\pi\int\limits_0^\infty dR\;R^2  \ e^{\hbox{$- \frac{\pi\left(R-{R}_c\right)^2}{\hbar G}$}}\;$ is the normalization factor. 

The wavefunction (\ref{solutionfin}) is an approximate wavefunction. When we evaluate expectation values to leading order in $C_{BH}$ it reproduces the correct answer up to subleading corrections. However, when we use it to evaluate some subleading corrections, the results reproduce the correct scaling but not necessarily the correct numerical factors or even signs.

An important observation about the average value $\langle R-R_c\rangle$  that we will need later is the following . Recall that we have required that the wavefunction reproduces the expectation value of the entropy  $\langle {S}_W  - A/ 4 l_p^2\rangle =0$. It follows that the average value of $R$ cannot be exactly $R_c$! It can only be equal to $R_c$ to leading order in $C_{BH}$. This is caused by the nonlinear dependence of $R$ on $S_W$, $\langle R  - R_c\rangle \sim \langle \sqrt{S_W}  - \sqrt{A/ 4 l_p^2}\rangle \ne 0$. The expansion of the square root necessarily involves higher orders of $R$ and at some point one of these higher order terms will have a non-vanishing expectation value. Generically, we expect a non-vanishing result already at the lowest possible order $\sim C_{BH}$ . However, the exact value of $\langle R  - R_c\rangle$ depends on subleading terms and therefore requires more detailed knowledge of the wavefunction. On the other hand, the expectation value $\langle (R  - R_c)^2\rangle$ can be evaluated to leading order and therefore its value $1/2 S_{BH}$ is robust against changes in the form of the wavefunction.

We will adhere to the form~(\ref{solutionfin})  for the purpose of keeping the presentation as simple as possible. This approximation  does not, however, change in a significant way any of our observations or conclusions. Our viewpoint is that the exact form of the wavefunction for the BH  is inconsequential to any of the basic findings.

The wavefunction should be interpreted in the semiclassical context of a fixed background. In our case, this is the classical Schwarzschild background. Further, a specific class of metrics that depends on a single parameter, $R$,  is assumed. Then the parameter of the metric becomes the relevant quantum variable, in what is the standard ``mini-superspace" approach. For this setup, one can use the background to define the Schwarzschild coordinates $\;r \ge R_c\;$ and treat the gravitational field as a gauge-fixed quantum variable in a fixed background.

An important lesson of the coupled-oscillator model is that a semiclassical regime leads to power-law corrections in its classicality parameter. For a BH, this parameter can similarly be read off  the exponent in the wavefunction
(also see Section~1) and goes as
\bea
C_{BH} \; = \;\frac{\hbar G}{\pi R_c^2}\;=\;\frac{1}{S_{BH}}\;,
\eea
which is  a very small (but still finite) number for a macroscopic  BH.

\subsection{Classical BHs as strong couplers for classical matter}

We begin here by reviewing how classical fields behave in the background of a classical BH. Let us now suppose that a  region in the proximity  of the horizon is being perturbed by massless scalar fields $\;\phi=\phi(t,r)\;$. After the source for the scalars has been turned off (say at $t=0$), the non-vanishing components of the  associated stress (energy-momentum) tensor go as
\be
T^{t}_{\;t}\;=\;  -\frac{1}{2}F^{-1}
\left[\nabla_t{\phi}\nabla_t{\phi} +\nabla_{r^{\ast}}\phi
\nabla_{r^{\ast}}\phi\right]\;,
\label{firstST}
\ee
\be
T^{r^{\ast}}_{\;r^{\ast}}\;=\; \frac{1}{2}F^{-1}
\left[\nabla_t{\phi}\nabla_t {\phi}+\nabla_{r^{\ast}}\phi
\nabla_{r^{\ast}}\phi\right]\;,
\ee
\be
T^{r^{\ast}}_{\;t}\;=\;-T^{t}_{\;r^{\ast}} \;=\;F^{-1}
\nabla_t{\phi}\nabla_{r^{\ast}}\phi\;,
\ee
\be
T^{\theta}_{\;\theta}\;=\; T^{\phi}_{\;\phi}\;=\;
\frac{1}{2}F^{-1}
\left[\nabla_t{\phi}\nabla_t {\phi}-\nabla_{r^{\ast}}\phi
\nabla_{r^{\ast}}\phi\right]\;,
\label{lastST}
\ee
where $\;F(r;R_c) \equiv -g_{tt} = \frac{r-R_c}{r}\;$
and $\;r^{\ast}=\int dr/F(r)\;$ is the usual Schwarzschild ``tortoise'' coordinate.

In the classical picture, the ``coupling''
\be
F^{-1} \;=\; \frac{r}{r-R_c}\;
\label{gtt}
\ee
blows up at the horizon, leading to a  divergent result.

However, we know from the no-hair theorem  that this cannot be the complete description. What really transpires is that the scalar
fields decay to zero magnitude in an exponentially short time. A simple way to understand the inevitability of this  rapid exponential decay is to view the scalars as small perturbations of the classical BH background. The behavior of the perturbations is then  determined by the  quasi-normal modes of the BH ({\it e.g.}, \cite{QNM} and the universality of exponential decay  is made particularly clear in \cite{QNM2}). After the fields have decayed, the strong coupling acts to freeze them; meaning that, after a very brief time, $\;T^{a}_{\ b}=0\;$.

This expectation can be put on firmer ground by calling upon Bekenstein's earliest proof of the no-hair theorem \cite{NHbek} (also see \cite{NHwin} and, in a slightly different context, \cite{hartle}).  We skip the initial steps and pick up the proof at what is, essentially, the integrated form of the radial-radial component of the field equation for the scalar fields,
\be
\left[F(r)r^2\phi\phi^{\prime}\right]^{\infty}_{R_c} \;=\;\int\limits^{\infty}_{R_c}dr\;r^2 F(r)\left(\phi^{\prime}\right)^2
\;,
\label{classJ}
\ee
where a prime denotes a differentiation by $r$ and the left-hand side consists
of boundary contributions that arise from an integration by parts.

The no-hair theorem comes about from the realization that the left-hand side vanishes because both of these surface contributions are vanishing. The $\;r=R_c\;$ contribution vanishes since $\;F=0\;$ at the horizon, where the fields are assumed to be bounded from above. The $r\to\infty$ contribution  vanishes because, by assumption, $\;\phi\to 0\;$ as $\;r\to\infty\;$.  The right-hand integrand, on the other hand, is manifestly non-negative. It is then clear that $\;\phi^{\prime}=0\;$ everywhere exterior to the horizon and, since $\phi$ already vanishes at infinity, $\phi=0$ follows.

\subsection{Classical fields in a semiclassical BH background}

But  what happens when the BH is treated semiclassically? This basic situation is similar to having an infinitely strong (classical) coupling between a pair of oscillators, and so we can look at our previous example for guidance. Now, once the BH  is treated as a quantum system, one can rather expect  via general arguments \cite{Dvali2} that no-hair theorems are violated. We find that this expectation is indeed realized  when allowing for a fluctuating background. Recalling the oscillator example, we will integrate out the BH fluctuations by using  the Gaussian wavefunction of Eq.~(\ref{solutionfin}).

The essential operator that we need to define to implement the proposed prescription is  $\wh{F}(r;R)$. The expectation value of this operator is the inverse of the classical coupling,
\be
\langle \psi_{BH}|  \wh{F}(r;R)  |\psi_{BH}\rangle\; =\; \frac{r-R}{r}\;.
\label{opeff}
\ee
The rule is then that we replace any appearance of $F(r)$ in a classical expression by the operator $\wh{F}(r;R)$ and evaluate expectation values using the Gaussian wavefunction of Eq.~(\ref{solutionfin}).

Following this procedure,  let us define a ``current''  operator $\wh{J}$
that corresponds to the horizon term  in the  left-hand side of Eq.~(\ref{classJ}),
\be
\wh{J}(r\to R_c;R)\;=\;\lim\limits_{r\to R_c}\;\wh{F}(r;R)\ r^2\ \phi(r) \phi^{\prime}(r)\;.
\ee

For the classical BH, the vanishing of $F(R_c)\ R_c^2\ \phi(R_c) \phi^{\prime}(R_c)$ via $\;F(R_c)=0\;$ was a crucial element in showing that $\phi$ vanishes everywhere. However, what now needs to be evaluated is
\be
\langle \psi_{BH}|  \wh{J}  |\psi_{BH}\rangle
 = 4\pi{\cal N}^{-1}\int\limits^{\infty}_0 dR\;R^2\frac{\left(R_c-R\right)}{R_c}
R_c^2\ \left[\phi\phi^{\prime}\right]_{r=R_c}\;
e^{-\frac{1}{C_{BH}}\frac{(R-R_c)^2}{R_c^2}}\;.
\label{previous}
\ee
The classical result is recovered when the limit $C_{BH}\to 0$ is taken before performing the integral. The square of the  wavefunction in this case becomes
$\delta(R-R_c)$ and $\langle \psi_{BH}|  \wh{J}  |\psi_{BH}\rangle$ vanishes.

As we have advocated, the correct prescription is that the integral needs to be performed while keeping  $C_{BH}$ finite. First consider that, for a massless field,  $\;\phi\sim 1/r\;$ and so  $\;\left[\phi\phi^{\prime}\right]_{r=R}\sim-\phi^2/R\;$. Otherwise, if the field is massive with mass $m$, then $\;\phi \sim e^{-mr}\;$ and $\;\left[\phi\phi^{\prime}\right]_{r=R}\sim - m \phi^2\;$.
In either case, the above integral can be rewritten as
\be
\langle \psi_{BH}|  \wh{J}  |\psi_{BH}\rangle
 \;=\; 4\pi{\cal N}^{-1}C_{BH}R^5_c\left[\phi\phi^{\prime}\right]_{r=R_c}
\int\limits^{\infty}_{-C_{BH}^{-1/2}}  dl\;\left(1+2 C^{1/2}_{BH}l+C_{BH}l^2\right) l \;
e^{-l^2}\;,
 \ee
where the polynomial in the round brackets is the result of expanding out
the factor of $R^2$ inside the integrand of Eq.~(\ref{previous}).

Now, as terms with odd powers of $l$ yield
exponentially small quantities
as per Eq.~(\ref{expdec}), we need only consider the term
quadratic in $l$. This leads to the outcome
\be
\langle \psi_{BH}|  \wh{J}  |\psi_{BH}\rangle
\;=\;
C_{BH}R^2_c\left[\phi\phi^{\prime}\right]^2_{r=R_c} \;,
\ee
where we have used
$\;\int dl\; l^2\; e^{-l^2}=\frac{1}{2}\int dl\; e^{-l^2}\;$ and
$\;{\cal N}\sim  4\pi R_c^2\sqrt{\pi C_{BH}R_c^2}\;$.
Recall that the average value of $\wh{J}$ is sensitive to the detailed form of the wavefunction, however, the fact that it does not vanish and is of order $C_{BH}$, is robust.

One could also consider the variance of $\wh{J}$ as it is more robust than the average of $\wh{J}$, depending only on the leading order form of the wavefunction. As we now show the variance of $\wh{J}$ is nonvanishing and of order $C_{BH}$, $\;\Delta J\sim C_{BH}\;$, rather than of order $C_{BH}^2$.
Since $\langle \wh{J} \rangle \sim C_{BH}$
$\;\Delta J^2 \sim \langle \wh{J}^2 \rangle\;$ and, in similar fashion to
the above calculation,
\bea
\langle \psi_{BH}|  \wh{J}^2  |\psi_{BH}\rangle
 &=& 4\pi{\cal N}^{-1}C^{3/2}_{BH}R^7_c\left[\phi\phi^{\prime}\right]^2_{r=R_c}
\int\limits^{\infty}_{-C_{BH}^{-1/2}}  dl\; l^2\;
e^{-l^2} \nonumber \\
&=& \frac{1}{2}C_{BH}R^4_c\left[\phi\phi^{\prime}\right]^2_{r=R_c}
+{\cal O}[C^2_{BH}]\;,
 \eea
where the lower line
follows from  $\;\int dl\; l^2\; e^{-l^2}=\frac{1}{2}\int dl\; e^{-l^2}\;$,
$\;{\cal N}\sim  4\pi R_c^2\sqrt{C_{BH}R_c^2}\;$ and the higher-order
terms are due to our approximated form in the top line.

This means that the average value of $\wh{J}$ and the variance of $\wh{J}$,  although suppressed by a  power of $C_{BH}$, can still  be measured during the lifetime of the BH. For a macroscopic BH, this suppression is indeed huge and  it means that, in practice, measuring the charge of the BH is an enormous challenge. But, as a matter of principle, it is no longer true that $\phi$ needs to vanish everywhere outside the horizon. Then, in principle, by measuring the average of $\wh{J}$ or its variance, one can determine the value of $\phi$ in the vicinity of the horizon. For a massless field, this can be done by evaluating the average  on a Gaussian surface away from the horizon. If the field is massive, this Gaussian surface has to be put at a distance less than $1/m$ from the horizon.  In conclusion, the value of $\phi$ or, equivalently, the global charge or Baryon number of the BH, is accessible in principle to an exterior observer.

One might then wonder about the quasi-normal mode perspective on this matter. However, the  usual exponential decay with time  results from fixing an ingoing boundary condition at the classical horizon. At the quantum (would-be) horizon, which is not a rigid one-way membrane like its classical counterpart, the fields no longer satisfy  this constraint. We elaborate on the transparency of the quantum horizon in
Subsection~3.4.

\subsection{Quantum fields in a semiclassical BH background}

When discussing quantum fields in a semiclassical BH background, one has to specify, in addition to the state of the BH, the state of the matter fields.
For current considerations, this is the  massless scalar field $\phi$,
and so the quantum state in question can be expressed as
\be
|\psi_{BH,\phi}\rangle \;=\;|\psi_{BH}\rangle\;|\psi_{\phi}\rangle
\;.
\ee

It is standard to assume that the matter is in its vacuum state outside the BH. This is a reasonable assumption when studying BHs that have had time to relax to an almost stationary configuration. But, then,  which vacuum state is the
appropriate one?

This choice of vacuum state has long been viewed as an important issue. An
asymptotic observer would choose a state that is devoid of ``Hawking-particle''
excitations, which means  the Boulware vacuum \cite{BoulV}. On the other hand,
for  a free-falling observer, the Unruh vacuum \cite{UnrV} is the correct
choice.~\footnote{This assumes that the BH  is slowly  evaporating away; for one in equilibrium with its surroundings, such as a large BH  in anti-de Sitter space, the correct choice is the closely related Hartle--Hawking vacuum \cite{HHV}.} The Unruh  vacuum  contains an infinite thermal bath of Hawking modes, and one is instructed to subtract off this contribution from the stress tensor.
The tensor is then rendered finite. Since, in our context, finite translates into weak coupling, the semiclassical treatment would not lead to fundamental changes.

Although there is a  significant difference in  energy between the two states, this discrepancy can   be compensated for with a sufficiently large boost or, equivalently, an appropriate Bogolubov transformation.  That this is true can be viewed as a consequence of the equivalence principle.

For instance, choosing the Bogolubov perspective
(see, {\it e.g.}, \cite{BD}), we can consider
a spacelike slice of the near-horizon geometry and write
\be
\phi(t,r) \;=\; \sum_{j} f_j(t,r) a_j  + f_j^{\ast}(t,r)a^{\dag}_j \;,
\ee
where  $j$ collectively labels the associated  quantum numbers,
 $a_j$ is an annihilation operator for the Unruh vacuum and the
$f$'s form a complete set of orthonormal, positive-frequency solutions.
The same field can also be written  as
\be
\phi(t,r)  \;=\; \sum_{k} g_k(t,r) b_k  + g_k^{\ast}(t,r)b^{\dag}_k \;,
\ee
where  $b_k$ is an annihilation operator for the Boulware vacuum
and $\;g_k=\sum_j\alpha_{jk}f_j -\beta_{jk} f_k^{\ast}\;$ such
that
the so-called Bogolubov coefficients
are required to satisfy
$\;\sum_j \alpha_{jk}\alpha^{\ast}_{jk^{\prime}}-\beta_{jk}\beta_{jk^{\prime}}^{\ast}
=\delta_{k,k^{\prime}}\;$.

The usual quandaries associated with the Hawking process can be traced to divergent quantities in the Boulware frame, as the Bogolubov transformation then becomes an ill-defined procedure. However, with the divergences censored, as expected when the BH is treated semiclassically, it is no longer an issue to transform from the Boulware frame to any other one. Indeed, the equivalence principle assures us that any two observers, even if one is free-falling through the horizon and the other is stationary, are in agreement on physically meaningful observables. A large stress-tensor component --- as long as it is finite --- has no particular meaning in this regard. The true physical observables are represented  by scalar quantities, and it is only these that the observers must agree on.

To see explicitly how the divergences are tamed for a finite $C_{BH}$,   we can evaluate a typical expectation value; for instance, that  of the
energy density
$\;\langle \psi_{BH,\phi}|  \wh{\rho}  |\psi_{BH,\phi}\rangle
= -\langle \psi_{BH,\phi}| T^t_{\;t}| \psi_{BH,\phi}\rangle\;$ at time $\;t=0\;$.

For classical fields, this is a straightforward calculation, but  quantum fields in a BH spacetime require a  more elaborate one. The calculation of an expectation value in  the state $|\psi_{\phi}\rangle$ requires knowing about the associated  density of the states. Fortunately, this is already known by virtue of `t Hooft's  famous brick-wall calculation \cite{hooftbw}. One can observe that  this method applies to energy density just as it does to entropy \cite{isrbw} and, moreover, the analysis of \cite{RJ} makes it clear that any reasonable smoothing function (in our case, the Gaussian for the BH) serves just as well as 't Hooft's hard cutoff.

The starting point is to consider the Klein--Gordon equation
for a massless  scalar field in Schwarzschild coordinates,
\begin{equation}
F^{-1}E^{2}\phi+ \frac{1}{r^2}\partial_{r}\left(r^2 F\partial_{r}\right)\phi -\frac{l\left(l+1\right)}{r^{2}}\phi=0\;,
\end{equation}
where $E$ is the energy and $l$ is the angular-momentum quantum number for a given mode. One  assumes a  WKB form for the scalar-field quantum  wavefunction $\;\psi_{\phi}\sim e^{\pm i\int dr\; p_r}\;$. This amounts to choosing the  Boulware vacuum, as evident from  the substantial red-shifting of wavelengths in this frame.

One  then  defines the   wavenumber
$p_r$ by
\begin{equation}
p_r^{2}= F^{-2} E^{2}-F^{-1}\frac{l\left(l+1\right)}{r^{2}}\;.
\end{equation}
This relation can  be
used  to  calculate the scalar field's  density of states $n$.
After the angular  parameter $l$ is integrated out, the result is
\cite{hooftbw,RJ}
\be
\frac{dn}{dr}\;=\; -\frac{2}{3\pi}E^3 r^2 F^{-2}(r)\;.
\label{dense}
\ee

Using this outcome, we show in Appendix~B that the near-horizon
energy density is given by
\be
\langle \psi_{\phi}|  \wh{\rho}  |\psi_{\phi}\rangle_{r\gtrsim R_c}
\;=\; -\frac{1}{7680\pi^2}\frac{1}{ \wh{F}^2(R_c;R) R_c^4}\;.
\label{energydense}
\ee
The negative sign is expected because the ground-state energy of the Boulware vacuum diverges to $-\infty$ at  a (classical) horizon. But what is important is that, for the semiclassical picture, typical matrix elements will go quadratically in the coupling $F^{-1}$  and not just linearly as they do for classical fields. The specific reason here is that the energy for a massless scalar field can be viewed as the variance of the momentum.

What is left is to compute the full expectation value, which means including and then integrating out the wavefunction for the black hole. Thus,
\be
\langle \psi_{BH,\phi}|  \wh{\rho}  |\psi_{BH,\phi}\rangle
\;=\; -\frac{1}{1920\pi}{\cal N}^{-1}\int\limits^{\infty}_0 dR\;R^2 \frac{1}{R_c^4}
\frac{R_c^2}{(R_c-R)^2}\;
e^{-\frac{1}{C_{BH}}\frac{(R-R_c)^2}{R_c^2}}\;.
\label{exact}
\ee
Again, the classical divergent result is recovered when one takes the limit $C_{BH}\to\infty$  before evaluating the integral.

Analogously to some earlier calculations,  Eq.~(\ref{exact}) can be well approximated by
\be
\langle \psi_{BH,\phi}|  \wh{\rho}  |\psi_{BH,\phi}\rangle
\;=\; -\frac{1}{1920\pi}{\cal N}^{-1}C^{-1/2}_{BH}R_c^{-1}\int\limits^{\infty}_{-C_{BH}^{-1/2}}dl\;
\frac{1}{l^2}\;
e^{-l^2}\;.
\label{approx}
\ee

We now recall that
$\;\int\limits_{-\infty}^{\infty} dl\; l^{-2} e^{-l^2}= 2\Gamma(-1/2)=- 4 \sqrt{\pi}\;$
and $\;{\cal N}= 4\pi R_c^2\sqrt{\pi C_{BH}R_c^2}\;$
(both up to insignificant deviations)
to arrive at
\be
\langle \psi_{BH,\phi}|  \wh{\rho}  |\psi_{BH,\phi}\rangle
\;=\; \frac{1}{1920\pi^2}\frac{1}{R_c^4}\frac{1}{C_{BH}}\;.
\label{approx2}
\ee
This is obviously a large (positive!) number but just as certainly a finite one.

One may observe that $\rho$ is still divergent if
extrapolated to the classical limit.
Yet, the qualitative pictures in the two regimes are
substantially  different.
Classically, there is an  infinite energy density at first, but
it  is decaying away  exponentially
fast via the  no-hair law.
 Semiclassically, on the other hand, there is a finite energy density that decays away only
by a power law.
But none of this should  come as a surprise because
a quantum theory, even when restricted to a domain
of semiclassicality, abhors both zeros and poles.

\subsection{Exponential versus power law suppression of matrix elements}

As now made evident, a semiclassical computation of the energy density and other typically computed matrix elements will lead to a power series expansion in terms of $\;C_{BH}= 1/S_{BH}\;$.
That is, something like
\be
\langle \rho \rangle_{r\to R_c} \;=\; \frac{1}{R_c^4}\left[b_{0}S_{BH}  + b_{1}\ln{S_{BH}} + b_2 +b_3S_{BH}^{-1} +b_4S_{BH}^{-2}+\dots
\right]\;,
\label{powser}
\ee
where the $b$'s are meant as  dimensionless numerical coefficients, some
of which could be vanishing.~\footnote{Although some coefficients
may accidentally vanish, further
corrections are  expected because of
the  approximations used in going from Eq.~(\ref{exact}) to
Eq.~(\ref{approx2}). That the series includes only integer powers of $S_{BH}^{-1}$ follows by analogy from
Eq.~(\ref{expdec}).}
The natural interpretation is  a renormalized version of what is a divergent classical result plus power-series corrections. One might rather have expected only an exponentially small correction $e^{-aS_{BH}}$ to a classical outcome; however, we can now see that some  matrix elements do not comply with this naive expectation.

One might then ask as to which matrix elements (if any) might  have only exponentially suppressed corrections. Following up on an argument made in \cite{RB}, we expect that this will be the case whenever one is probing sufficiently deep into the BH  interior. For instance, suppose that an external observer wants to measure
the charge density $\rho_Q$ of the scalar fields at a radius of $\;r_{\epsilon}<R_c\;$, as discussed in the previous section. Classically, the result is of course zero, as the external observer will be unable to retrieve any probe once it has passed through the $\;r=R_c\;$ causal barrier. Determining what transpires semiclassically is, on the other hand, tantamount to asking as to what extent does  the relevant part of the interior
become transparent. More formally, this question can be  rephrased as ``What is the probability ${\cal P}_{\epsilon}$ that the quantum horizon takes on a value $\;R\leq r_{\epsilon}\;$ in any given measurement?"

It follows from our previous discussion that the probability ${\cal P}_{\epsilon}$ is given by
\bea
{\cal P}_{\epsilon} & = & \frac{4\pi}{{\cal N}}
\int\limits^{r_{\epsilon}}_0 dR\;R^2\; e^{-\frac{1}{C_{BH}} \frac{(R-R_c)^2}{R^2_c}}
  \sim  \left(\int\limits^{R_c}_{0} dR\;   e^{-\frac{1}{C_{BH}} \frac{(R-R_c)^2}{R^2_c}}\right)^{-1}
\int\limits^{r_{\epsilon}}_0 dR\;  e^{-\frac{1}{C_{BH}} \frac{(R-R_c)^2}{R^2_c}}
 \nonumber \\
&\sim & e^{- \frac{1}{C_{BH}} \frac{(R_c-r_{\epsilon})^2}{R_c^2}}\;.
\eea
Now, recalling that $\;C_{BH}R_c^2=\hbar G/\pi\sim l_p^2\;$ and assuming a separation from the horizon $R_c-r_{\epsilon}$ that is not too small in Planck units,
one obtains  ${\cal P}_{\epsilon} \sim e^{-aS}$. Under the same circumstance, $\wh{F}^{-2}$ (which is relevant to the density of states) becomes a dimensionless number of order unity and the density  can then be expected to go as
\be
\langle\rho_Q\rangle_{r=r_{\epsilon}}\;\sim\; R_c^{-4}e^{-aS}\;,
\ee
which is  an exponentially suppressed correction to the classical zero, as claimed. Notice that, at least in this case, ``deep into the interior'' really just means a distance that is parametrically larger than a Planck length.

There is another way to understand the distinction between power law and exponential corrections \cite{Malsr}. If modes with small wavelengths $\;\lambda\ll R_c\;$  could contribute to the BH  radiative process, this would likely spoil the thermal spectrum that  was not only predicted by Hawking \cite{Hawk}, but later substantiated   by independent string-theory calculations \cite{MalStro}. Consequently,  significant deviations from the classical picture can only be tolerated for long-wavelength modes $\;\lambda\sim R_c\;$, with the additional provision that the deviations are small enough so as to keep the thermal spectrum (approximately) intact. This is completely  consistent with our picture, whereby the non-trivial, modest, power-law corrections are   restricted to within only  a few Planck lengths from the horizon.

\subsection{Firewalls}

Let us now address  the firewall paradox that
was recently proposed by Almheiri, Marolf, Polchinski and Sully \cite{AMPS}. (See \cite{fw1,fw2,fw3,fw4,avery,lowe,vv,pap} for a sample of the subsequent debate.) For current purposes, the basic argument of \cite{AMPS} can be phrased as follows \cite{avery}: If the evaporation of a  BH  is to be a unitary process, then the  quantum state describing the vicinity of the horizon  must store information about the collapsing matter system, as it is this information that is supposedly carried off by  the outgoing Hawking modes. On the other hand, if a free-falling observer is to see nothing special on her way through the horizon, then the Unruh vacuum is the obligatory quantum state. Anything else and the observer would  encounter a sea of high-energy quanta; that is,  a firewall. However, the Unruh vacuum is independent of the matter that formed the BH  and, as such, incapable of storing the requisite information. Herein lies the paradox.

But our findings from Subsection~3.3  circumvent  this issue by making it clear that the choice of quantum state is inconsequential. Put simply, a fluctuating background can be expected to tame the infinitely large energies associated with non-trivial quantum states; meaning that information about the collapsing matter can be safely  stored after all. How this stored information gets transferred into the outgoing radiation is another story that  will be told in an upcoming article \cite{infoBM}.

As for the strong-subadditivity-of-entropy version of the firewall paradox (see any of the cited papers for an explanation), we agree with \cite{pap} that the resolution is because the information stored in the various subsystems (early radiation, late radiation, in-falling partners of late radiation) cannot be cleanly separated in the way that the argument implies. However, what \cite{pap} did miss is that a true classical horizon --- which acts as a rigid barrier between regions of spacetime --- would  in fact allow for such a separation. What saves the day is a quantum (fluctuating) horizon for which it is impossible, even  in principle, to say what spacetime region  the entanglement is ``living''. As has been known since the famous EPR thought experiment, quantum entanglements are fundamentally non-local.

\section{Conclusion}

We now summarize the main results
\begin{itemize}
\item
We have proposed that BHs have to be treated as fluctuating quantum objects and that it suffices to treat them to leading order in the classicality parameter $\;C_{BH}= 1/S_{BH}\;$. To evaluate expectation values of matter operators in a BH background, it is necessary to include and then integrate out  the BH fluctuations prior to calculating the  matter matrix elements.

\item
We have found that, when the BH is treated semiclassically as directed above,
classically forbidden processes are no longer censored. They are rather suppressed by powers of the classicality parameter $C_{BH}$. For example, a global charge of the BH can now be measured, in principle.  This point can be understood intuitively; even if the back-reaction of the matter field fluctuations on the background is small, that of the background fluctuations on the fields need not be when the coupling is strong.

\item
We have found that, when the BH is treated semiclassically as discussed above
{\em and} the matter is treated quantum mechanically, then divergent quantities, such as the energy density at the horizon, are rendered finite and proportional to inverse powers of the classicality parameter $1/C_{BH}$. This point can also be understood intuitively;  a fluctuating surface of classically infinite redshift can no longer have  an infinite redshift.

\item
The previous results offer a simple resolution to longstanding conceptual issues like the existence of global symmetries in the presence of BHs, the trans-Planckian problem, as well as the more recent firewall puzzle. The information  paradox is also resolved by similar arguments, as will be discussed in a following article \cite{infoBM}.
\end{itemize}

Before concluding, let us briefly elaborate on the last point. One can consider one of two situations when addressing such conceptual issues. Either it is the classical situation of an infinitely massive BH  or else the semiclassical regime when the mass is a large but finite. For an infinitely massive BH, the situation is simple. For instance, the concern over information loss or firewalls is rendered irrelevant, as an infinitely massive BH  never evaporates. The classical
BH rather acts as an infinite reservoir of Hawking particles in the same way that the Earth acts as an infinite source or sink for electrons in electrical circuits.

As for the trans-Planckian problem,
any mode with an exponentially large frequency  will still be parametrically less energetic than the BH  itself.  On the other hand, a  mode with an infinite frequency is one that must have passed through the horizon and then can already
be considered part of the (infinitely massive)  black hole interior.

Then what about the semiclassical case?  There can still be none of the claimed issues, although the reasoning is now  different. There is no longer a true (classical) singularity through which information can be lost nor  a horizon for which the Hawking modes could be red-shifted without bound. Of course, their would-be surrogates still represent regions of high curvature and very large redshift, respectively. But, inasmuch as there are no longer any exact infinities nor any causally inaccessible regions of spacetime, there is also no good  reason to believe that such a situation could give rise to paradoxical physics. Indeed, without any infinities, an evaporating BH  cannot be conceptually any different than the proverbial burning book.

In our opinion, the real paradox has been  in trying to merge both classical and semiclassical elements into the same set of thought experiments. One is then guilty of trying to retain  her ``conceptual cake'' while consuming it too.

\section*{Acknowledgments}

The research of RB was supported by the Israel Science Foundation grant no. 239/10. The research of AJMM received support from
a Rhodes University Discretionary Grant RD35/2012.

\appendix
\section{More on harmonic oscillators}

Let us now address the issue of  having neglected the interaction term $\frac{1}{2}(\wh{k}-k_c)\wh{x}^2$
when determining the state function~(\ref{wavey}).
Clearly, the interaction term is a small quantity, having an exponentially
suppressed  expectation value in any stationary state. However,
such a term can still be expected to perturb  $|\psi\rangle$ away from its form in Eq.~(\ref{wavey})  and
thereby lead to  additional corrections to the energy. Nevertheless, such corrections
are found to be inconsequential when compared to the leading perturbative order in Eq.~(\ref{correct}).

To understand this last point, let us call upon standard quantum perturbative theory
(see, {\it e.g.}, \cite{sakarai}) and consider an initial (unperturbed) Hamiltonian $\wh{H_0}$
with eigenstates $\left\{|L^{(0)}\rangle\right\}$ and associated eigenvalues $\left\{E_L^{(0)}\right\}$.
If we add a perturbation $\wh{V}$ to the Hamiltonian, then
the leading-order correction to the eigenstate  $|N^{(0)}\rangle$  goes as
\be
\delta |N^{(1)}\rangle\;=\; \sum_{M\neq N} \frac{\langle M^{(0)} | \wh{V} | N^{(0)} \rangle}{E_N^{(0)}- E_M^{(0)}} \;,
\label{wfcorrect}
\ee
whereas the leading-order correction to its energy goes as
\be
\delta E_N^{(1)} \;=\; \langle N^{(0)}| \wh{V} | N^{(0)} \rangle\;.
\ee
Notice that only the unperturbed states and energies are ever used in
the calculation.

In our present case, $\wh {V}$ is the interaction term $\frac{1}{2}(\wh{k}-k_c)\wh{x}^2$, $\wh{H}_0$ is
the remainder of the Hamiltonian~(\ref{hammy}) and, importantly,
$|N^{(0)}\rangle$ includes (irrespective of the $x$-part of the wavefunction)
the ground state of the background oscillator.
Hence,  any non-vanishing matrix element $\langle K^{(0)} |\wh{V}| L^{(0)} \rangle$ must involve at least
one  excited state of
the background oscillator, as  $\;\langle N^{(0)} |\wh{k}-k_c|  N^{(0)}\rangle=0\;$
up to exponentially small corrections ({\em cf}, Eq.~(\ref{expdec})).
For this reason, the first-order energy shift is  vanishingly small and  we must instead call upon the second,
\be
\delta E_N^{(2)} \;=\;\sum_{M\neq N} \frac{\left|\langle M^{(0)} | \wh{V} | N^{(0)} \rangle\right|^2}
{E_N^{(0)}- E_M^{(0)}} \;.
\ee

Because of the orthogonality of  oscillator states, the  constant $k_c$ cannot
contribute to any of  these matrix elements.
Consequently, the correction of interest can be written  (with the superscripts  implied) as
\be
\delta E \;=\; \frac{|\langle \wh{x}^2 \rangle |^2}{2}\sum_{p\neq 0}
\frac{\left|\langle p_k  | \wh{k} | 0_k \rangle\right|^2}{\hbar{\cal C}
(0-p)} \nonumber \\
\;=\;
 -\frac{|\langle \wh{x}^2 \rangle |^2}{2}
\frac{\left|\langle 1_k  | \wh{k} | 0_k \rangle\right|^2}{\hbar{\cal C}}
\;,
\ee
where we have used that the unperturbed wavefunction factorizes to get the factor
$|\langle \wh{x}^2 \rangle |^2$. The second equality follows from the fact that the ``position'' operator
$\wh{k}$ can only mix eigenstates that are one number apart
and the denominators follow from the identification of ${\cal C}^2$ as the spring constant for the background oscillator.

Now, referring to the standard quantum-oscillator formalism and keeping in mind that ${\cal C}^2$ is the heavy oscillator's spring constant, one can promptly deduce that
$\;\left|\langle \wh{x}^2 \rangle\right|^2 \sim  \hbar^2/k_c\;$ and $\;\left|\langle 1_k | \wh{k}| 0_k \rangle\right|^2
\sim \hbar/{\cal C}\;$. Hence, the net result is that $\;\delta E/E \sim C_{OSC}\ (k_c^{1/2}/{\cal C})\;$, which is suppressed by a factor of $\;k_c^{1/2}/{\cal C}=\omega_x/\omega_k \ll 1\;$ relative to the leading correction in Eq.~(\ref{correct2}).

By the same reasoning, the  leading correction to the wavefunction~(\ref{wfcorrect}) goes as
$\;\hbar^{1/2}/(k_c^{1/2}{\cal C}^{3/2})\sim C_{OSC}^{1/2}\ k_c^{1/2}/{\cal C}\;$ and so is similarly suppressed.

\section{Energy density of quantum fields}

Here, we substantiate Eq.~(\ref{energydense}) for the expectation value
$\;\rho_{\phi}\equiv \langle \psi_{\phi}|  \wh{\rho}  |\psi_{\phi}\rangle\;$.
For this calculation, $\;\hbar=1\;$.

Let us begin with a defining relation for the free energy $F_{\phi}$ of a
thermal bath of
scalar bosons,
\be
e^{-\beta F_{\phi}}\;=\;\prod_{n, l, m} \left(1-e^{-\beta E_{n,l}}\right)^{-1}\;,
\ee
where $n$ is the principal quantum number, $l$ and $m$ are the usual angular-momentum
quantum numbers, $\beta^{-1}$ is the temperature of the bath and $E_{n,l}$ is the energy
of a given level. Solving for $F_{\phi}$ and taking the continuum limit, we have
\be
F_{\phi}\;=\;\frac{1}{\beta}\int dl\;(2l+1)\int dn \; \ln{\left(1-e^{-\beta E(n,l)}
\right)}\;
\ee
or, after integrating by parts,
\be
F_{\phi}\;=\; -\int dl\;(2l+1)\int dE \; \frac{1}{e^{\beta E}-1} n(E,l) \;.
\ee
This last form makes it clear that $n$ functions as the density of states.

Let us next recall Eq.~(\ref{dense}),
\be
\frac{dn}{dr}\; =\; -\frac{2}{3\pi}E^3 r^2 F^{-2}(r)\;,
\label{dense2}
\ee
from the main text. As this expression is  obtained
after already integrating out the angular quantum numbers,
we can dismiss the above integral over $l$.  Eq.~(\ref{dense2}) also
indicates  that $n$ scales
as $E^3$, and so we can define an energy-independent density
 $\widetilde{n}$ via the relation $\;n(E)=E^3\widetilde{n}\;$.
Consequently,
\be
F_{\phi} \;=\;
-\widetilde{n}\int\limits^{\infty}_{0} dE \; \frac{E^3}{e^{\beta E}-1}
\;=\; -\widetilde{n} \frac{\pi^4}{15\beta^4}
\;.
\ee

The energy of the scalars is then given by $E_{\phi}=\frac{\partial(\beta F_{\phi})}{\partial\beta}\;$
or
\be
E_{\phi} \;=\; \widetilde{n}\frac{\pi^4}{5\beta^4}
\;=\; \int dr\;\frac{d\widetilde{n}}{dr}\frac{\pi^4}{5\beta^4}\;.
\ee
This can be compared to~\footnote{Although a spatial integral, there is no factor of $\sqrt{g_{rr}}$ appearing
in the integrated form of the energy \cite{MTW}.}
\be
E_{\phi} \;=\; \int dr  \; 4\pi r^2\rho_{\phi}\;,
\ee
from which it can be deduced (with the help of Eq.~(\ref{dense2})) that
\be
\rho_{\phi}\;=\; \frac{\pi^3}{20\beta^4}r^{-2}\frac{d\widetilde{n}}{dr}\nonumber\\
\;=\; -\frac{\pi^2}{30\beta^4}F^{-2}(r)\;.
\ee

Finally, for an asymptotic observer probing near the horizon, the appropriate
value of $\beta$ is fixed by the inverse of the Hawking temperature,
$\;\beta=1/T_H= 4\pi R_c\;$.
Hence, we arrive at the claimed result
\be
\rho_{\phi}(r\gtrsim R_c)\;=\;-\frac{F^{-2}(r)}{7680\pi^2 R_c^4}\;.
\ee


\begin{thebibliography}{99}

\bibitem{BD}  N. D. Birrell and P. C. W. Davies, {\em Quantum fields in curved space}
(Cambridge University Press, Cambridge, 1982).

\bibitem{SF} S. A. Fulling, {\em  Aspects of quantum field theory in curved space-time}
(Cambridge University Press, Cambridge, 1982).

\bibitem{RW} R. M. Wald, {\em Quantum field theory in curved space-time and black hole thermodynamics}
(University of Chicago Press, Chicago, 1995).





\bibitem{Hawk} S. W. Hawking, ``Black hole explosions'',  Nature {\bf 248}, 30  (1974); ``Particle creation
by black holes'',
Comm. Math. Phys. {\bf 43}, 199 (1975).


\bibitem{RB}
  R.~Brustein,
  ``Origin of the blackhole information paradox,''
  arXiv:1209.2686 [hep-th].
  %%CITATION = ARXIV:1209.2686;%%



\bibitem{Dvali1}
  G.~Dvali and C.~Gomez,
  ``Black Hole's Quantum N-Portrait,''
  arXiv:1112.3359 [hep-th];
  %%CITATION = ARXIV:1112.3359;%%
``Black Hole's 1/N Hair,''
  arXiv:1203.6575 [hep-th];
  %%CITATION = ARXIV:1203.6575;%%
  ``Black Holes as Critical Point of Quantum Phase Transition,''
   arXiv:1207.4059 [hep-th];
  ``Black Hole Macro-Quantumness,''
  arXiv:1212.0765 [hep-th].
  %%CITATION = ARXIV:1212.0765;%%


%\cite{Veneziano:2012yj}
\bibitem{Veneziano}
  G.~Veneziano,
  ``Quantum hair and the string-black hole correspondence,''  arXiv:1212.2606 [hep-th].  
  %%CITATION = ARXIV:1212.2606;%%  




\bibitem{RM} R.~Brustein and M.~Hadad,
  ``Wave function of the quantum black hole,''
  arXiv:1202.5273 [hep-th].
  %%CITATION = ARXIV:1202.5273;%%


\bibitem{englert}
F.~Englert and P.~Spindel,
  ``The Hidden horizon and black hole unitarity,''
  JHEP {\bf 1012}, 065 (2010)
  [arXiv:1009.6190 [hep-th]].
  %%CITATION = ARXIV:1009.6190;%%



\bibitem{FB1}
S.~Giusto and S.~D.~Mathur,
  ``Fuzzball geometries and higher derivative corrections for extremal holes,''
  Nucl.\ Phys.\ B {\bf 738}, 48 (2006)
  [hep-th/0412133].
  %%CITATION = HEP-TH/0412133;%%


\bibitem{FB2}
  S.~D.~Mathur,
  ``The Fuzzball proposal for black holes: An Elementary review,''
  Fortsch.\ Phys.\  {\bf 53}, 793 (2005)
  [hep-th/0502050];
  ``The Quantum structure of black holes,''
  Class.\ Quant.\ Grav.\  {\bf 23}, R115 (2006)
  [hep-th/0510180];
S.~D.~Mathur,
  ``Black Holes and Beyond,''
  arXiv:1205.0776 [hep-th].
  %%CITATION = ARXIV:1205.0776;%%
  %%CITATION = HEP-TH/0510180;%%
  %%CITATION = HEP-TH/0502050;%%








\bibitem{info1}
  S.~W.~Hawking,
  ``Black Holes and Unpredictability,''
  Phys.\ Bull.\  {\bf 29}, 23 (1978).
  %%CITATION = PHSBB,29,23;%%

\bibitem{info2}
  D.~N.~Page,
  ``Black hole information,''
  hep-th/9305040.


\bibitem{info3}
  S.~B.~Giddings,
  ``Comments on information loss and remnants,''
  Phys.\ Rev.\ D {\bf 49}, 4078 (1994)
  [hep-th/9310101].
  %%CITATION = HEP-TH/9310101;%%



\bibitem{info4}
  S.~D.~Mathur,
  ``What Exactly is the Information Paradox?,''
  Lect.\ Notes Phys.\  {\bf 769}, 3 (2009)
  [arXiv:0803.2030 [hep-th]];
  %%CITATION = ARXIV:0803.2030;%%
``The Information paradox: A Pedagogical introduction,''
  Class.\ Quant.\ Grav.\  {\bf 26}, 224001 (2009)
  [arXiv:0909.1038 [hep-th]];
  %%CITATION = ARXIV:0909.1038;%%
 ``What the information paradox is {\it not},''
  arXiv:1108.0302 [hep-th].
  %%CITATION = ARXIV:1108.0302;%%




\bibitem{trans1} G. W. Gibbons, ``Quantum processes near black holes``,
in {\em Proceedings of the Marcel Grossman meeting on Recent
Advances in the Fundamentals of General Relativity}, ed. R. Ruffini,
449 (North Holland, Amsterdam, 1977).


\bibitem{trans2}
T.~Jacobson,
``Black hole evaporation and ultrashort distances,''
Phys.\ Rev.\ D {\bf 44}, 1731 (1991);
  ``Black hole radiation in the presence of a short distance cutoff,''
  Phys.\ Rev.\ D {\bf 48}, 728 (1993)
  [hep-th/9303103];
 %%CITATION = HEP-TH/9303103;%%




\bibitem{trans3} A.~D.~Helfer,
  ``Do black holes radiate?,''
  Rept.\ Prog.\ Phys.\  {\bf 66}, 943 (2003)
  [gr-qc/0304042].
  %%CITATION = GR-QC/0304042;%%



\bibitem{AMPS}
A.~Almheiri, D.~Marolf, J.~Polchinski and J.~Sully,
  ``Black Holes: Complementarity or Firewalls?,''
  arXiv:1207.3123 [hep-th].
  %%CITATION = ARXIV:1207.3123;%%



\bibitem{RJ}
R.~Brustein and J.~Kupferman,
  ``Black hole entropy divergence and the uncertainty principle,''
  Phys.\ Rev.\ D {\bf 83}, 124014 (2011)
  [arXiv:1010.4157 [hep-th]].
  %%CITATION = ARXIV:1010.4157;%%


\bibitem{Dvali2}
  G.~Dvali, C.~Gomez and D.~Lust,
  ``Black Hole Quantum Mechanics in the Presence of Species,''
  arXiv:1206.2365 [hep-th].
  %%CITATION = ARXIV:1206.2365;%%



\bibitem{CL}  	
	A. O. Caldeira and A. J.  Leggett,
``Path integral approach to quantum Brownian motion,''
	Physica A {\bf 121}, 587 (1983).



\bibitem{fluc1} J. W. York, ``Dynamical origin of black-hole radiance,''
Phys. Rev. D {\bf 28}, 2929 (1983).

\bibitem{fluc2}
  L.~H.~Ford and N.~F.~Svaiter,
  ``Cosmological and black hole horizon fluctuations,''
  Phys.\ Rev.\ D {\bf 56}, 2226 (1997)
  [gr-qc/9704050];
  %%CITATION = GR-QC/9704050;%%
  ``Vacuum energy density near fluctuating boundaries,''
  Phys.\ Rev.\ D {\bf 58}, 065007 (1998)
  [quant-ph/9804056].
  %%CITATION = QUANT-PH/9804056;%%




\bibitem{fluc3}
C.~Barrabes, V.~P.~Frolov and R.~Parentani,
  ``Stochastically fluctuating black hole geometry, Hawking radiation and the transPlanckian problem,''
  Phys.\ Rev.\ D {\bf 62}, 044020 (2000)
  [gr-qc/0001102].
  %%CITATION = GR-QC/0001102;%%


\bibitem{fluc4}
  R.~Parentani,
  ``Quantum metric fluctuations and Hawking radiation,''
  Phys.\ Rev.\ D {\bf 63}, 041503 (2001)
  [gr-qc/0009011].
  %%CITATION = GR-QC/0009011;%%


\bibitem{fluc5} R.~T.~Thompson and L.~H.~Ford
  ``Enhanced Black Hole Horizon Fluctuations,''
  Phys.\ Rev.\ D {\bf 78}, 024014 (2008)
  [arXiv:0803.1980 [gr-qc]].
  %%CITATION = ARXIV:0803.1980;%%








\bibitem{RV} S. M. Roy  and A. Venugopalan,
``Exact Solutions of the Caldeira-Leggett Master Equation: A Factorization
Theorem for Decoherence,''
 arXiv:quant-ph/9910004.



\bibitem{QNM}
K.~D.~Kokkotas and B.~G.~Schmidt,
  ``Quasinormal modes of stars and black holes,''
  Living Rev.\ Rel.\  {\bf 2}, 2 (1999)
  [gr-qc/9909058].
  %%CITATION = GR-QC/9909058;%%

\bibitem{QNM2}
A.~J.~M.~Medved, D.~Martin and M.~Visser,
  ``Dirty black holes: Quasinormal modes,''
  Class.\ Quant.\ Grav.\  {\bf 21}, 1393 (2004)
  [gr-qc/0310009].
  %%CITATION = GR-QC/0310009;%%



\bibitem{NHbek} J. D. Bekenstein, ``Nonexistence of baryon number for static black holes,"
Phys. Rev. D {\bf 5}, 1239 (1972).


\bibitem{NHwin}
  E.~Winstanley,
  ``On the existence of conformally coupled scalar field hair for black holes in (anti-)de Sitter space,''
  Found.\ Phys.\  {\bf 33}, 111 (2003)
  [gr-qc/0205092].
  %%CITATION = GR-QC/0205092;%%


%\cite{Hartle:1971qq}
\bibitem{hartle}
  J.~B.~Hartle,
  ``Long-range neutrino forces exerted by kerr black holes,''  Phys.\ Rev.\ D {\bf 3}, 2938 (1971).
  %%CITATION = PHRVA,D3,2938;%%



\bibitem{BoulV}
D. G. Boulware, ``Quantum field theory in Schwarzschild and Rindler spaces'',
Phys. Rev. D {\bf 11}, 1404 (1975).



\bibitem{UnrV}
  W.~G.~Unruh,
  ``Notes on black hole evaporation,''
  Phys.\ Rev.\ D {\bf 14}, 870 (1976).


\bibitem{HHV}
J.~B.~Hartle and S.~W.~Hawking,
  ``Path Integral Derivation of Black Hole Radiance,''
  Phys.\ Rev.\ D {\bf 13}, 2188 (1976).
  %%CITATION = PHRVA,D13,2188;%%




\bibitem{hooftbw} G. 't Hooft,
``On the quantum structure of a black hole,'',
 Nucl. Phys. B {\bf 256}, 727 (1985).

\bibitem{isrbw} S.~Mukohyama and W.~Israel,
  ``Black holes, brick walls and the Boulware state,''
  Phys.\ Rev.\ D {\bf 58}, 104005 (1998)
  [gr-qc/9806012].
  %%CITATION =







\bibitem{Malsr} J.~M.~Maldacena, ``The black hole information problem'',
a lecture presented  at the conference {\em Forty Years of Black
Hole Thermodynamics} in Jerusalem, Israel (2012).


\bibitem{MalStro}
  J.~M.~Maldacena and A.~Strominger,
  ``Black hole grey body factors and d-brane spectroscopy,''
  Phys.\ Rev.\ D {\bf 55}, 861 (1997)
  [hep-th/9609026].
  %%CITATION = HEP-TH/9609026;%%




\bibitem{fw1}
L.~Susskind,
  ``Singularities, Firewalls, and Complementarity,''
  arXiv:1208.3445 [hep-th];
  ``The Transfer of Entanglement: The Case for Firewalls,''
  arXiv:1210.2098 [hep-th].
  %%CITATION = ARXIV:1210.2098;%%

 \bibitem{fw2}
  R.~Bousso,
  ``Complementarity Is Not Enough,''
  arXiv:1207.5192 [hep-th].
  %%CITATION = ARXIV:1207.5192;%%

\bibitem{fw3}
  Y.~Nomura, J.~Varela and S.~J.~Weinberg,
  ``Complementarity Endures: No Firewall for an Infalling Observer,''
  arXiv:1207.6626 [hep-th];
  ``Black Holes, Information, and Hilbert Space for Quantum Gravity,''
  arXiv:1210.6348 [hep-th].
  %%CITATION = ARXIV:1210.6348;%%

  %%CITATION = ARXIV:1207.6626;%%

\bibitem{fw4}
  S.~D.~Mathur and D.~Turton,
  ``Comments on black holes I: The possibility of complementarity,''
  arXiv:1208.2005 [hep-th].
  %%CITATION = ARXIV:1208.2005;%%


\bibitem{avery}
  S.~G.~Avery, B.~D.~Chowdhury and A.~Puhm,
  ``Unitarity and fuzzball complementarity: 'Alice fuzzes but may not even know it!',''
  arXiv:1210.6996 [hep-th].
  %%CITATION = ARXIV:1210.6996;%%

\bibitem{lowe}
  K.~Larjo, D.~A.~Lowe and L.~Thorlacius,
  ``Black holes without firewalls,''
  arXiv:1211.4620 [hep-th].
  %%CITATION = ARXIV:1211.4620;%%


\bibitem{vv}  E.~Verlinde and H.~Verlinde,
  ``Black Hole Entanglement and Quantum Error Correction,''
  arXiv:1211.6913 [hep-th].
  %%CITATION = ARXIV:1211.6913;%%

\bibitem{pap}
  K.~Papadodimas and S.~Raju,
  ``An Infalling Observer in AdS/CFT,''
  arXiv:1211.6767 [hep-th].
  %%CITATION = ARXIV:1211.6767;%%


\bibitem{infoBM} R. Brustein and A. J. M. Medved, work in progress.






\bibitem{sakarai} J. J. Sakurai, {\em Modern Quantum Mechanics (Revised edition)}
(Addison-Wesley, Reading, Massachusetts, 1994).



\bibitem{MTW}
C. W. Misner, K. S. Thorne and J. A. Wheeler,
{\em Gravitation} (W. H. Freeman, San
Francisco, 1973).


\end{thebibliography}
\end{document}